\documentclass[twocolumn]{aastex701}

\shorttitle{TransFit-CSM}
\shortauthors{Zhang et al.}
\graphicspath{{./}{figures/}}
\usepackage{amsmath}
\usepackage{threeparttable}
\usepackage{booktabs}
\usepackage{changepage} 

\graphicspath{{./}{figures/}}

\begin{document}

\title{ \texttt{TransFit-CSM}: A Fast, Physically Consistent Framework for Interaction-Powered Transients}

\author[0009-0000-2423-6825]{Yu-Hao Zhang}
\affiliation{Institute of Astrophysics, Central China Normal University, Wuhan 430079, China; \url{liuld@ccnu.edu.cn;yuyw@ccnu.edu.cn}}
\affiliation{Education Research and Application Center, National Astronomical Data Center, Wuhan 430079, China}
\affiliation{Key Laboratory of Quark and Lepton Physics (Central China Normal University), Ministry of Education, Wuhan 430079, China}
\email{zhang-yh@mails.ccnu.edu.cn}

\author[0000-0002-8708-0597]{Liang-Duan Liu}
\affiliation{Institute of Astrophysics, Central China Normal University, Wuhan 430079, China; \url{liuld@ccnu.edu.cn;yuyw@ccnu.edu.cn}}
\affiliation{Education Research and Application Center, National Astronomical Data Center, Wuhan 430079, China}
\affiliation{Key Laboratory of Quark and Lepton Physics (Central China Normal University), Ministry of Education, Wuhan 430079, China}
\email{liuld@ccnu.edu.cn}

\author{Ze-Xin Du}
\affiliation{Institute of Astrophysics, Central China Normal University, Wuhan 430079, China; \url{liuld@ccnu.edu.cn;yuyw@ccnu.edu.cn}}
\affiliation{Education Research and Application Center, National Astronomical Data Center, Wuhan 430079, China}
\affiliation{Key Laboratory of Quark and Lepton Physics (Central China Normal University), Ministry of Education, Wuhan 430079, China}
\email{duzexin@mails.ccnu.edu.cn}

\author{Guang-Lei Wu}
\affiliation{Institute of Astrophysics, Central China Normal University, Wuhan 430079, China; \url{liuld@ccnu.edu.cn;yuyw@ccnu.edu.cn}}
\affiliation{Education Research and Application Center, National Astronomical Data Center, Wuhan 430079, China}
\affiliation{Key Laboratory of Quark and Lepton Physics (Central China Normal University), Ministry of Education, Wuhan 430079, China}
\email{wuguanglei@mails.ccnu.edu.cn}

\author[0009-0004-9719-272X]{Jing-Yao Li}
\affiliation{Institute of Astrophysics, Central China Normal University, Wuhan 430079, China; \url{liuld@ccnu.edu.cn;yuyw@ccnu.edu.cn}}
\affiliation{Education Research and Application Center, National Astronomical Data Center, Wuhan 430079, China}
\affiliation{Key Laboratory of Quark and Lepton Physics (Central China Normal University), Ministry of Education, Wuhan 430079, China}
\email{lijy@mails.ccnu.edu.cn}

\author[0000-0002-1067-1911]{Yun-Wei Yu}
\affiliation{Institute of Astrophysics, Central China Normal University, Wuhan 430079, China; \url{liuld@ccnu.edu.cn;yuyw@ccnu.edu.cn}}
\affiliation{Education Research and Application Center, National Astronomical Data Center, Wuhan 430079, China}
\affiliation{Key Laboratory of Quark and Lepton Physics (Central China Normal University), Ministry of Education, Wuhan 430079, China}
\email{yuyw@ccnu.edu.cn}

\begin{abstract}
We introduce \texttt{TransFit-CSM}, a fast and physically consistent framework for modeling interaction-powered transients. The method self-consistently couples the ejecta–circumstellar-medium (CSM) shock dynamics to radiative diffusion from a moving heating boundary that is tied to the shocks. In this way, both the photon escape path and the effective diffusion time evolve with radius and time. We numerically solve the mass–momentum equations for the forward and reverse shocks together with the diffusion equation in the unshocked CSM. As a result, \texttt{TransFit-CSM} reproduces the canonical sequence of an early dark phase, a diffusion-mediated rise and peak, and a post-interaction cooling tail, and it clarifies why Arnett-like peak rules break down in optically thick CSM. The framework is Bayesian-ready and constrains physical parameters of the ejecta and CSM from bolometric or joint multi-band light curves. Applications to SN~2006gy and SN~2010jl demonstrate accurate fits and physically interpretable posteriors. These fits highlight the dominant role of pre-supernova mass loss in shaping the observables. Because it is both computationally efficient and physically grounded, \texttt{TransFit-CSM} bridges simple analytic prescriptions and radiation-hydrodynamic simulations. This capability enables population-level inference for current and upcoming time-domain surveys.

\end{abstract}

\keywords{ Supernovae (1668);  Circumstellar matter (241);  Stellar mass loss (1613); Core-collapse supernovae (304)}

\section{Introduction} \label{sec:intro}

A supernova (SN) marks the violent terminal explosion of a massive star. Throughout their lifetimes, such stars drive strong winds that create a dense CSM. When the SN explodes, the rapidly expanding ejecta collides with this CSM, generating powerful shock waves. These shocks heat the gas and produce X-ray emission, a fraction of which can be reprocessed into optical and ultraviolet radiation. In extreme cases, this reprocessed emission can dominate the intrinsic luminosity of the SN ejecta itself. Indeed, some Type~IIn SNe are mainly powered by this interaction, where the kinetic energy of the ejecta is efficiently converted into radiation (for reviews, see \citealt{Dessart2024}).

Modeling SN ejecta and CSM interaction is essential for probing the mass-loss history of massive stars in the years to decades before core collapse. The physics of such an interaction has been extensively studied (e.g. \citealt{Chevalier1982, Chevalier1994,Dessart2015, Margalit2022,Khatami2024}), with both analytical and numerical work predicting the resulting light curves and spectra (e.g. \citealt{Chevalier2011, Ginzburg2012, Morozova2017}) and applications to specific events (e.g. \citealt{Ofek2010,Liu2018a,Leung2020,Hu2025}).

Detailed radiation-hydrodynamic simulations provide an accurate description of shock dynamics and radiative transfer. Numerical investigations of this process began several decades ago (e.g., \citealt{Falk1977}), with modern state-of-the-art calculations now performed using codes such as \texttt{STELLA} (\citealt{Blinnikov1993, Blinnikov1998}). These simulations are crucial for understanding light-curve and spectral formation in interacting supernovae (e.g., \citealt{vanMarle2010, Moriya2013, Dessart2015, Takei2024}), but their high computational cost renders them prohibitive for applications like Bayesian inference across large SN samples.

To enable efficient parameter estimation, simplified semi-analytic models are commonly adopted, most notably the framework of \citet{Chatzopoulos2012, Chatzopoulos2013} based on Arnett’s \citeyearpar{Arnett1980,Arnett1982} diffusion formalism. While computationally tractable, such treatments assume centrally deposited heating and approximate the forward and reverse shocks as independent sources, often producing inconsistent luminosity predictions. Comparisons with radiation-hydrodynamic simulations confirm that these simplifications can bias estimates of CSM properties (e.g., \citealt{Sorokina2016}). Despite these limitations, such analytic models are integrated into widely used software packages like \texttt{MOSFiT} \citep{Nicholl2017,Guillochon2018} and \texttt{Redback} \citep{Sarin2024}, and are extensively applied to fit lightcurves of interaction-powered transients.

The key difficulty lies in analytically treating radiative diffusion through a medium where moving shocks are simultaneously depositing energy. As both the forward shock (FS) and reverse shock (RS) propagate, they continually modify the effective optical depth and photon escape time. This coupling turns the energy equation into a non-separable partial differential equation with diffusion terms \citep{Chatzopoulos2012,Moriya2018}.

In our previous work, we developed \texttt{TransFit}, a modular framework for fast and flexible modeling of SN light curves powered by radioactive decay, shock-deposited internal energy, and magnetar central engines \citep{Liu2025}. \texttt{TransFit} provides Bayesian inference capabilities via Markov Chain Monte Carlo (MCMC) sampling while retaining close agreement with detailed radiative-transfer calculations. Importantly, that framework assumes energy sources whose spatial position is fixed in the comoving frame (e.g., a central magnetar or radioactive heating approximated by a fixed mass coordinate). Building on this foundation, we present \texttt{TransFit-CSM}, which extends the framework to interaction-powered transients. It self-consistently couples thin-shell ejecta–CSM dynamics with a shock-tied, moving energy-injection boundary and a time-dependent diffusion domain. This removes the limitation of fixed heating locations in our previous model, captures how outward shocks reduce the photon escape path, and provides a computationally efficient yet physically consistent description of ejecta–CSM interaction.

This paper is organized as follows. Section~\ref{sec:framework} outlines the \texttt{TransFit-CSM} framework---coupled thin-shell shock dynamics with radiative diffusion from a moving, shock-tied heating boundary. Section~\ref{sec:Lightcurve} explores the resulting light-curve morphologies, particularly the distinction between compact and extended CSM. Section~\ref{sec:Applications} applies \texttt{TransFit-CSM}  to well-observed interacting SNe (e.g., SN~2006gy and SN~2010jl). Finally, Section~\ref{sec:Conclusions} concludes with a summary of our main results and an outlook for future work.

\begin{figure*}
    \centering
\includegraphics[width=0.9\textwidth]{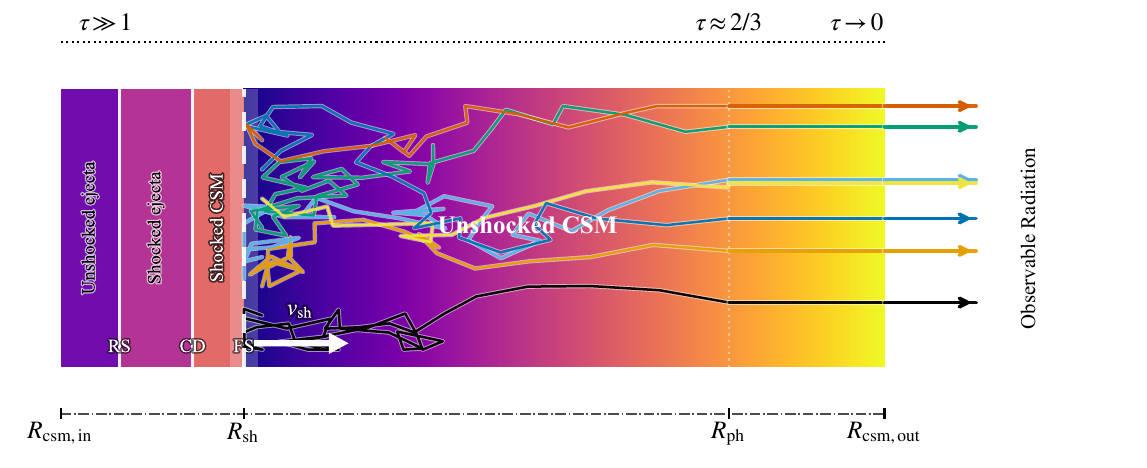}
\caption{Schematic illustration of the SN--CSM interaction within the \texttt{TransFit-CSM} framework. Photons generated at the shock front ($R_{\rm sh}$) are initially trapped in the optically thick medium ($\tau \gg 1$) and diffuse outward through multiple scatterings (random-walk trajectories). Once they reach the photosphere ($R_{\rm ph}$, where $\tau \approx 2/3$), the medium becomes optically thin, and photons free-stream toward a distant observer. Key radii and the shock velocity ($v_{\rm sh}$) are indicated for clarity.}
    \label{Fig:Schem}
\end{figure*}

\section{Physical Framework} \label{sec:framework}

When a massive star explodes as a SN, the ejecta expand at velocities of order \(10^{4}\,\mathrm{km\,s^{-1}}\). If the progenitor experienced substantial pre-explosion mass loss, a dense CSM surrounds the star.  This collision establishes a pair of shocks structure: a FS propagates outward into the CSM, while a RS travels back into the ejecta, compressing and heating the gas in both regions. In this process, the kinetic energy of the ejecta is efficiently converted into thermal energy at the shocks, which is then reprocessed into the radiation that powers the transient event.

As schematically illustrated in Figure~\ref{Fig:Schem}, photons generated at the shock front, located at radius \(R_{\mathrm{sh}}\), are initially trapped within an optically thick medium where \(\tau \gg 1\). These photons subsequently propagate outward through the unshocked CSM via radiative diffusion, a process characterized by multiple scattering events (shown as random-walk trajectories). Observable radiation emerges once the photons reach the photosphere at radius \(R_{\mathrm{ph}}\), the location where the optical depth to the observer decreases to \(\tau \approx 2/3\). At this point, the medium becomes optically thin, allowing the photons to free-stream toward the observer. The dynamical evolution and radiative transport of this coupled system ultimately determine the morphology of the observed light curve, which is dictated by the physical properties of both the ejecta and the CSM---a direct reflection of the progenitor's evolution and mass-loss history.

\begin{figure}
    \centering
    \includegraphics[width=1\linewidth]{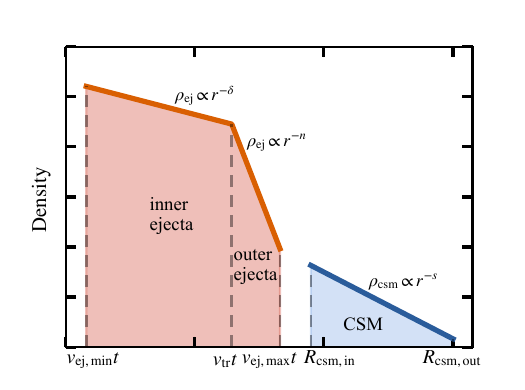}
    \caption{Default density structure of the supernova ejecta and CSM adopted in \texttt{TransFit-CSM}. The ejecta profile is represented by a broken power law, with a flat inner core ($\rho_{\mathrm{ej}}\propto r^{-\delta}$) and a steep outer envelope ($\rho_{\mathrm{ej}}\propto r^{-n}$). The transition velocity $v_{\mathrm{tr}}$ marks the boundary between the inner and outer ejecta components. The CSM is modeled as a power-law density distribution ($\rho_{\mathrm{csm}}\propto r^{-s}$), extending from $R_{\mathrm{csm,in}}$ to $R_{\mathrm{csm,out}}$.  }
    \label{fig:CSMSNdensity}
\end{figure}

\subsection{Initial conditions}

In this subsection, we specify the initial conditions adopted in \texttt{TransFit-CSM}. The default density structures of the SN ejecta and the CSM are shown in Figure~\ref{fig:CSMSNdensity}. The system comprises freely expanding ejecta inside a low-density gap and a dense CSM outside. We take the initial ejecta outer radius to be the progenitor’s stellar radius \(R_\star\), and denote the inner edge of the dense CSM by \(R_{\rm csm,in}\) with \(R_{\rm csm,in}>R_\star\). Because of this gap, the ejecta do not interact with the CSM immediately. The arrival time of the fastest ejecta at the CSM is
\begin{equation}
t_{\rm in} \;=\; \frac{R_{\rm csm,in}-R_\star}{v_{\rm ej,max}}
\;\approx\; \frac{R_{\rm csm,in}}{v_{\rm ej,max}}
\quad (R_\star \ll R_{\rm csm,in}),
\end{equation}
where \(v_{\rm ej,max}\) is the outermost ejecta velocity implied by the ejecta kinetic energy and density profile. For \(t<t_{\rm in}\), the evolution is internal to the ejecta (e.g., shock cooling and/or radioactive heating). At \(t\ge t_{\rm in}\), a FS forms at \(r\simeq R_{\rm csm,in}\) and propagates into the CSM, while a RS travels back into the ejecta.

\subsubsection{Structure of SN ejecta}

Following shock breakout, the ejecta quickly relax to homologous expansion, such that the velocity is proportional to radius, $v\simeq r/t$. It is therefore convenient to adopt velocity as the coordinate and to factor out the universal $t^{-3}$ decline of density, writing
\begin{equation} \label{eq:rho_homologous}
  \rho_{\mathrm{ej}}(r,t)=\rho_{\mathrm{ej,in}}
  \left(\frac{t}{t_{\mathrm{in}}}\right)^{-3}\eta_{\mathrm{ej}}(v),
\end{equation}
where $\rho_{\mathrm{ej,in}}$ is a characteristic density scale and $\eta_{\mathrm{ej}}(v)$ specifies the normalized velocity profile.

For core-collapse SNe, $\eta_{\mathrm{ej}}(v)$ is often represented by a broken power law that captures the flat inner ``core'' and the steep outer ``envelope'':
\begin{equation} \label{eq:broken_power_law}
  \eta_{\mathrm{ej}}(v)=
  \begin{cases}
    \left(\dfrac{v}{v_{\mathrm{tr}}}\right)^{-\delta}, & v_{\mathrm{ej,min}}\le v < v_{\mathrm{tr}}, \\[8pt]
    \left(\dfrac{v}{v_{\mathrm{tr}}}\right)^{-n}, & v_{\mathrm{tr}}\le v \le v_{\mathrm{ej,max}},
  \end{cases}
\end{equation}
where $v_{\mathrm{tr}}$ is the transition velocity and $\delta$ and $n$ are the inner and outer slopes, respectively. Finiteness of the total ejecta mass and kinetic energy requires $n>5$ and $\delta<3$. Typical values are $\delta\simeq 0$--$1$; $n\simeq 10$ for stripped-envelope progenitors (Types~Ib/Ic), and $n\simeq 12$ for red-supergiant explosions \citep{Matzner1999}.

Alternative profiles are motivated by different progenitors.  SNe~Ia ejecta are commonly approximated by an exponential velocity distribution \citep{Nomoto1984,Dwarkadas1998},
\begin{equation} \label{eq:exp_profile}
  \eta_{\mathrm{ej}}(v)\propto \exp\!\left(-\frac{v}{v_{\rm sc}}\right),
\end{equation}
where $v_{\rm sc}$ is a characteristic velocity scale. In all cases, the normalization of $\eta_{\mathrm{ej}}(v)$ together with $\rho_{\mathrm{ej,in}}$ is chosen such that the integrated density and energy recover the total ejecta mass and kinetic energy,
\begin{equation}
  \int 4\pi r^2 \rho_{\mathrm{ej}}\,\mathrm{d}r  = M_{\mathrm{ej}}, 
  \qquad 
  \int 4\pi r^2 \left(\tfrac{1}{2}\rho_{\mathrm{ej}} v^2\right)\,\mathrm{d}r = E_{\mathrm{sn}}.
\end{equation}

\subsubsection{Structure of CSM}

CSM can originate from a variety of processes that operate during the late evolution of massive stars. Episodic or eruptive outbursts, such as those observed in luminous blue variables or driven by late-stage nuclear burning instabilities, can expel large amounts of mass into the immediate vicinity of the progenitor (e.g., \citealt{ Quataert2012,Smith2014, Fuller2017}). In addition, binary interactions—including Roche-lobe overflow, common-envelope evolution, and mergers—are efficient at removing the H-rich envelope and creating complex, often asymmetric, CSM structures (\citealt{Podsiadlowski1992, Yoon2017}). Finally, steady stellar winds, ranging from the slow dense winds of red supergiants to the fast winds of Wolf–Rayet stars, contribute significantly to shaping CSM over long timescales (e.g., \citealt{Vink2001, Beasor2020}). This diversity in both the origin and geometry of the CSM naturally leads to a wide range of transient light-curve morphologies observed in interacting SNe (e.g., \citealt{Chevalier1994, Moriya2013}).

We model the CSM as a shell that extends from an inner radius $R_{\mathrm{csm,in}}$ to an outer radius $R_{\mathrm{csm,out}}$. The density profile is given by
\begin{equation}
    \rho_{\mathrm{csm}}(r) = \rho_{\mathrm{csm,in}} \eta_{\mathrm{csm}}(r),
\end{equation}
where $\rho_{\mathrm{csm,in}}$ is the density at the inner boundary and $\eta_{\mathrm{csm}}(r)$ is a dimensionless profile normalized such that $\eta_{\mathrm{csm}}(R_{\mathrm{csm,in}}) = 1$. For our model, we adopt the commonly used power-law profile
\begin{equation}
\eta_{\mathrm{csm}}(r)=\left(\frac{r}{R_{\mathrm{csm,in}}}\right)^{-s},
\end{equation}
with $s$ as the power-law index. The canonical case of a steady, spherically symmetric stellar wind with a constant mass-loss rate ($\dot{M}$) and terminal velocity ($v_w$) corresponds to a density profile of $\rho_{\mathrm{csm}}(r) = \dot{M} / (4\pi r^2 v_w)$, which implies $s=2$. Departures from this value ($s \neq 2$) reflect different mass-loss histories. For example, recent work suggests that the long-rising light curves of some interaction-powered transients, which can take hundreds of days to reach peak brightness, may be explained by a flatter CSM density profile with $s < 1.5$ \citep{Moriya2023}.

While this power-law framework provides a useful approximation, observations of interacting supernovae reveal significant structural complexities. The CSM is often not a smooth, spherically symmetric envelope but can exhibit substantial inhomogeneities. For example, evidence for clumpiness has been inferred from the temporal evolution of narrow emission lines in SNe IIn, suggesting that the interaction occurs with dense, discrete structures rather than a uniform medium \citep{Smith2009}.

\subsection{Shock dynamics}

In the presence of efficient radiative cooling, thermal energy is rapidly removed from the post-shock region. This causes the pressure-supported layer between the forward and reverse shocks to collapse into a geometrically thin, dense shell. Consequently, the standard energy-conserving self-similar solutions, which assume an adiabatic shock and resolve the finite shell width \citep{Chevalier1982, Nadyozhin1985}, are not applicable. Instead, in the limit where the cooling length is much smaller than the shock radius, the system's evolution is governed by mass and momentum conservation alone \citep{Moriya2013}.

We therefore model the shocked region as a single thin shell with mass $M_{\mathrm{sh}}$, radius $R_{\mathrm{sh}}$, and velocity $v_{\mathrm{sh}}$. The shell's equation of motion is determined by balancing the ram pressures from the unshocked ejecta and CSM \citep{Chevalier1982,Moriya2013}:
\begin{equation} \label{Eq:Momentum}
M_{\mathrm{sh}} \frac{\mathrm{d} v_{\mathrm{sh}}}{\mathrm{d} t}
= 4\pi R_{\mathrm{sh}}^{2}\!\left[
\rho_{\mathrm{ej}}\,\bigl(v_{\mathrm{ej}} - v_{\mathrm{sh}}\bigr)^{2}
- \rho_{\mathrm{csm}}\,\bigl(v_{\mathrm{sh}} - v_{\mathrm{csm}}\bigr)^{2}
\right],
\end{equation}
where $v_{\mathrm{ej}} = R_{\mathrm{sh}}/t$ is the ejecta velocity at the shell radius (assuming homologous expansion), $v_{\mathrm{sh}} = \mathrm{d}R_{\mathrm{sh}}/\mathrm{d}t$, and $v_{\mathrm{csm}}$ is the CSM speed. The rate of mass accumulation in the shell is given by \citep{Chevalier1982,Moriya2013}:
\begin{equation} \label{Eq:Mass}
\frac{\mathrm{d} M_{\mathrm{sh}}}{\mathrm{d} t}
= 4\pi R_{\mathrm{sh}}^{2}\!\left[
\rho_{\mathrm{ej}}\,\bigl(v_{\mathrm{ej}} - v_{\mathrm{sh}}\bigr)
+ \rho_{\mathrm{csm}}\,\bigl(v_{\mathrm{sh}} - v_{\mathrm{csm}}\bigr)
\right],
\end{equation}
where $M_{\mathrm{sh}}$ is the total swept-up mass of both ejecta and CSM.

For power-law density profiles in the outer ejecta ($\rho_{\mathrm{ej}} \propto r^{-n}$) and the CSM ($\rho_{\mathrm{csm}} \propto r^{-s}$), these momentum-conserving equations yield a self-similar solution after an initial transient phase \citep{Chevalier1982}. The shell radius then evolves as a power law in time:
\begin{equation}
R_{\mathrm{sh}} (t) \propto t^{m}, \qquad m = \frac{n - 3}{n - s}.
\end{equation}
This solution applies once the shell has expanded far beyond its initial radius ($R_{\mathrm{sh}} \gg R_{\mathrm{csm,in}}$), such that the initial conditions no longer influence the dynamics.

While this self-similar framework is widely applied to the analysis of interacting supernovae, its validity is restricted to cases where both the ejecta and CSM density profiles can be described by power laws. The solution breaks down, for example, if the shock propagates into the inner, flatter-density regions of the ejecta. Therefore, instead of adopting the self-similar solution, we perform our calculations by numerically solving the thin-shell equations for momentum conservation Eq.(\ref{Eq:Momentum}) and mass conservation Eq.(\ref{Eq:Mass}). This direct numerical approach allows us to determine the temporal evolution of the shell's radius ($R_{\mathrm{sh}}$), velocity ($v_{\mathrm{sh}}$), and mass ($M_{\mathrm{sh}}$) for more general density profiles. We numerically solve these dynamical equations using a non-dimensionalized scheme based on characteristic values, as detailed in Appendix~\ref{app:dimensionless}.

\subsection{Shock Power and Energy Deposition}\label{sec:shock_power}
In the thin-shell approximation, the observable luminosity arises from the dissipation of kinetic energy at the FS and RS. The power processed by each shock is set by the kinetic-energy flux through its front, \(\frac{1}{2}\rho v_{\mathrm{rel}}^3\), where \(v_{\mathrm{rel}}\) is the relative velocity between the shock and the upstream gas. The corresponding power contributions from FS and RS are:
\begin{align}
    L_{\mathrm{FS}}(t) &= 2\pi R_{\mathrm{sh}}^{2} \rho_{\mathrm{csm}}(R_{\mathrm{sh}}) \bigl(v_{\mathrm{sh}}-v_{\mathrm{csm}}\bigr)^{3}, \label{eq:Lfs} \\
    L_{\mathrm{RS}}(t) &= 2\pi R_{\mathrm{sh}}^{2} \rho_{\mathrm{ej}}(R_{\mathrm{sh}}) \bigl(v_{\mathrm{ej}}-v_{\mathrm{sh}}\bigr)^{3}, \label{eq:Lrs}
\end{align}
where \(R_{\mathrm{sh}}\) is the shock radius, \(\rho_{\mathrm{csm}}(R_{\mathrm{sh}})\) and \(\rho_{\mathrm{ej}}(R_{\mathrm{sh}})\) denote the upstream densities immediately ahead of the FS and RS, respectively, and \(v_{\mathrm{sh}}\), \(v_{\mathrm{csm}}\), and \(v_{\mathrm{ej}}\) are the velocities of the shock, CSM, and ejecta at that radius. For homologously expanding ejecta, the local ejecta velocity is \(v_{\mathrm{ej}}(R,t) \approx R/t\).

The total instantaneous power generated by the shocks is the sum of both contributions:
\begin{equation}
    L_{\mathrm{sh}}(t) = L_{\mathrm{FS}}(t) + L_{\mathrm{RS}}(t). \label{eq:Lsh}
\end{equation}
In reality, only part of this shock power is converted into thermal radiation. A significant fraction may be channeled into nonthermal processes, such as particle acceleration or high-energy emission. To account for this uncertainty, we introduce a thermalization efficiency factor $\epsilon_{\rm int}$, such that
\begin{equation}
L_{\mathrm{heat}}(t) = \epsilon_{\rm int} L_{\mathrm{sh}}(t).
\label{eq:Lheat}
\end{equation}

In our framework, $\epsilon_{\rm int}$ is treated as a free parameter, encapsulating the uncertain microphysics of kinetic-to-thermal energy conversion at the shocks. This approach allows the model to remain flexible and to capture the diversity of observed interaction-powered transients, while deferring the detailed physics of shock dissipation and radiative efficiency to future high-resolution radiation-hydrodynamic studies. Importantly, heating operates only during the interaction phase, and ceases once the shocks exit the CSM, after which the system evolves purely under radiative cooling of the stored thermal energy.

\subsection{Radiative Diffusion in the Unshocked CSM}\label{subsec:diffusion}

In traditional energy sources of SNe, such as radioactive decay or powering by a central engine, the heat source is typically assumed to be stationary at a specific location within the ejecta. By contrast, the source of shock heating is dynamic; it propagates outward with the shock front, continuously updating the local photon diffusion time.  

In the unshocked CSM, the evolution of the radiation energy is governed primarily by photon diffusion. Because $v_{\rm csm} \ll v_{\rm sh}$, the CSM is effectively stationary; we therefore track only energy transport through the unshocked layers and neglect bulk-motion and adiabatic-expansion terms.

\begin{equation}
  \frac{\partial E(r,t)}{\partial t}
  \;=\;
  -\,\frac{\partial L}{\partial m}
  \;+\;
  \dot{\epsilon}_{\rm sh}(r,t),
\end{equation}
where $L(r,t)$ is the luminosity as a function of radius coordinate $r$, and $\dot{\epsilon}_{\rm sh}(r,t)$ is the specific shock-heating rate.

\paragraph{Radiative cooling by diffusion}
In the diffusion approximation, the divergence of the radiative flux per unit mass is
\begin{equation}
  -\,\frac{\partial L}{\partial m}
  \;=\;
  \frac{1}{4\pi r^{2}\rho_{\rm csm}}
  \frac{\partial}{\partial r}
  \!\left[
     4\pi r^{2}\,
     \frac{c}{3\,\kappa\,\rho_{\rm csm}}\,
     \frac{\partial u}{\partial r}
  \right],
\end{equation}
where $\rho_{\rm csm}(r)$ is the local CSM density, and $u(r,t)=\rho_{\rm csm}(r) E(r,t)$ is the radiation energy density,  $\kappa$ is the opacity of unshocked CSM.

\paragraph{Shock heating and inner boundary condition.}
The source term $\dot{\epsilon}_{\rm sh}(r,t)$ represents the power input per unit 
mass from shock dissipation. Within the interaction framework, heating is confined 
to a geometrically thin region immediately adjacent to the shock front, with no 
deposition in the unshocked interior. Formally, this localized injection can be 
written as
\begin{equation}
  \dot{\epsilon}_{\rm sh}(r,t)
  \;=\;
  \dot{\epsilon}_{\rm sh,0}\,f_{\rm sh}(t)\,
  \delta\!\left[r - R_{\rm sh}(t)\right],
\end{equation}
where $f_{\rm sh}(t)$ describes the temporal evolution of the shock power and the 
Dirac delta function enforces localization at the instantaneous shock position 
$R_{\rm sh}(t)$. Here $\dot{\epsilon}_{\rm sh,0}$ carries units of 
$\mathrm{erg\,s^{-1}\,g^{-1}}$ to preserve dimensional consistency.

In practice, the delta-function source is implemented as an inner boundary condition 
at the base of the unshocked CSM. The flux through this boundary is set equal to the 
instantaneous shock luminosity per unit area,
\begin{equation}
  F(R_{\rm sh},t) =-\frac{c}{3\kappa\rho_{\rm csm}}
  \left.\frac{\partial u}{\partial r}\right|_{R_{\rm sh}}
  = \frac{L_{\rm heat}(t)}{4\pi R_{\rm sh}^2}.
\end{equation}
This Neumann-type boundary condition ensures 
that the total shock power $L_{\rm heat}(t)$ is consistently injected into the 
diffusion domain from below, thereby providing a practical numerical realization 
of the delta-function heating term. As the shock front advances, the inner boundary 
is updated accordingly, guaranteeing that energy deposition always occurs at the 
correct spatial location.

\paragraph{Outer boundary condition.}
To describe radiation escape at the outer boundary of the CSM, we adopt the gray Eddington approximation for a diffusive atmosphere, which relates the local temperature $T$ to the effective temperature $T_{\rm eff}$ of the emitting surface:
\begin{equation}
    T^{4}(\tau) \;=\; \frac{3}{4}\,T_{\rm eff}^{4}\,\Bigl(\tau+\frac{2}{3}\Bigr),
    \label{eq:eddington}
\end{equation}
where $\tau$ is the optical depth measured outward toward the observer. We impose this as the outer boundary condition at $r = R_{\rm ph}$, where $\tau=2/3$, yielding
\begin{equation}
    T^{4}(R_{\rm ph},t)=T_{\rm eff}^{4}(t).
    \label{eq:bc_outer_T}
\end{equation}
During the interaction phase, the bolometric luminosity is set by the emergent flux at the CSM photosphere,
\begin{equation}
L_{\rm bol}(t) = 4\pi \sigma R_{\rm ph}^{2} T^{4}(R_{\rm ph},t),
\end{equation}
where $\sigma$ is the Stefan–Boltzmann constant. The observed light curve therefore mainly traces the temporal evolution of the photospheric temperature.

\subsection{Shock-Cooling Phase}
\label{subsec:shockcooling}

The shock-heating phase ends once the shocks have no further material to traverse. 
The RS contribution, $L_{\mathrm{RS}}$, terminates when the RS has fully processed 
the ejecta and reaches its inner boundary. Similarly, the FS contribution, 
$L_{\mathrm{FS}}$, is truncated when the FS exits the CSM, i.e., when 
$R_{\mathrm{sh}} \geq R_{\mathrm{csm,out}}$. Beyond this point, no additional shock heating 
occurs, and the system enters a post-interaction shock-cooling phase dominated by the thermal 
relaxation of previously heated material.  

At the moment when the FS reaches the outer edge of the CSM, the circumstellar shell is no 
longer stationary but instead begins to expand together with the ejecta. Much like the ejecta 
itself in the aftermath of core collapse, the shocked CSM is blown outward at high velocities.

In this regime, we neglect any additional radioactive heating or central-engine power, and 
consider only the residual energy deposited by the shocks. Compared to the interaction stage, 
the governing equation now includes an adiabatic cooling term but lacks a heating source term. 
The evolution of the specific internal energy therefore follows
\begin{equation}
  \frac{\partial E}{\partial t} \;+\; P\,\frac{dV}{dt}
  \;=\; -\,\frac{\partial L}{\partial m},
\end{equation}
where $P$ is the local pressure and $V$ is the specific volume. The first term accounts for 
the decline of internal energy, the second describes adiabatic expansion losses as the shocked 
layers accelerate outward, and the third term captures radiative diffusion losses through 
the overlying material.  

Physically, the shock-cooling phase thus represents a passive leakage process, powered solely by the release of stored internal energy as the system undergoes adiabatic expansion. The emergent luminosity \(L(t)\) declines monotonically as this internal energy \(E\) decreases, producing a cooling tail analogous to that seen in the post-shock breakout phases of extended stellar envelopes (e.g., \citealt{Piro2015}). Simultaneously, the spectral energy distribution shifts to lower temperatures as the effective photosphere recedes and the shocked CSM expands.

\section{Light-Curve Morphologies: Compact vs. Extended CSM}\label{sec:Lightcurve}

In \texttt{TransFit-CSM}, we self-consistently calculate the coupled dynamics and radiative transfer during both the interaction and subsequent cooling phases. This approach tracks the diffusion of internal energy from the shock-heated layers and its effect on the surface temperature at the photosphere. To improve computational efficiency, we recast the governing equations into a dimensionless form; the full derivation and boundary conditions are given in Appendix~\ref{app:dimensionless}.

The morphology of interaction-powered transients is set by the competition between shock heating and radiative diffusion. 
In the diffusion approximation with isotropic scattering, the local 
``diffusion drift velocity'' can be written as
\begin{equation}
  v_{\mathrm{diff}}(r) \simeq \frac{c}{\tau(r)} .
\end{equation}
The mean time for photons to diffuse from the shock radius 
$R_{\mathrm{sh}}$ to the photospheric radius $R_{\mathrm{ph}}$ is
\begin{equation}
  t_{\mathrm{diff}}(r) \simeq 
  \int_{R_{\mathrm{sh}}}^{R_{\mathrm{ph}}} 
  \frac{\mathrm{d}r}{v_{\mathrm{diff}}(r)} 
  = \frac{1}{c}\int_{R_{\mathrm{sh}}}^{R_{\mathrm{ph}}} \tau(r)\,\mathrm{d}r .
\end{equation}

Here the optical depth depends on the CSM density profile 
$\eta_{\mathrm{csm}}(r)$ and the opacity $\kappa$, 
\begin{equation}
  \tau(r) = \int_r^{R_{\mathrm{csm,out}}} 
  \kappa\,\rho_{\mathrm{csm}}(r')\,\mathrm{d}r' ,
\end{equation}
and the local diffusion timescale can be expressed as
\begin{equation}
  t_{\mathrm{diff}}(r) = 
  \frac{\tau_0 R_{\mathrm{csm,out}}}{c}\,
  \Theta\!\left(r, R_{\mathrm{ph}}, R_{\mathrm{csm,out}}, 
  \eta_{\mathrm{csm}}\right),
\end{equation}
where 
$\tau_0 \equiv \kappa M_{\mathrm{csm}}/(4\pi R_{\mathrm{csm,out}}^2)$ 
defines a characteristic optical depth, and $\Theta$ is a 
dimensionless geometrical factor that accounts for the effects of the 
density profile and the shock position. 
The explicit form of $\Theta$ is
\begin{equation}
\Theta \equiv  
  \frac{R_{\mathrm{csm,out}}
  \int^{R_{\mathrm{ph}}}_{r_{\mathrm{sh}}} \int_{r}^{R_{\mathrm{csm,out}}}
  \eta_{\mathrm{csm}}(r')\,\mathrm{d}r'\,\mathrm{d}r}{
  \int^{R_{\mathrm{csm,out}}}_{R_{\mathrm{csm,in}}} 
  r^2\,\eta_{\mathrm{csm}}(r)\,\mathrm{d}r}.
\end{equation}
For a steady wind profile ($s=2$),
\begin{equation}
   \Theta \approx 
   \ln \frac{R_{\mathrm{ph}}}{r_{\mathrm{sh}}}
   - \frac{R_{\mathrm{ph}} - r_{\mathrm{sh}}}{R_{\mathrm{csm,out}}},
\end{equation}
and for a uniform-density CSM,
\begin{equation}
    \Theta \approx 
    \frac{3 (R_{\mathrm{ph}} - r_{\mathrm{sh}})}{R_{\mathrm{csm,out}}}
    \left[ 1 - \frac{R_{\mathrm{ph}} + r_{\mathrm{sh}}}
    {2 R_{\mathrm{csm,out}}}\right].
\end{equation}

These limiting cases have been widely discussed in the literature 
\citep[e.g.,][]{Chevalier2011,Ginzburg2012,Ginzburg2014,Wasserman2025}. 
Here we generalize the treatment to arbitrary density slopes and shock 
positions. 

\begin{figure}[h]
    \centering
    \includegraphics[width=1\linewidth]{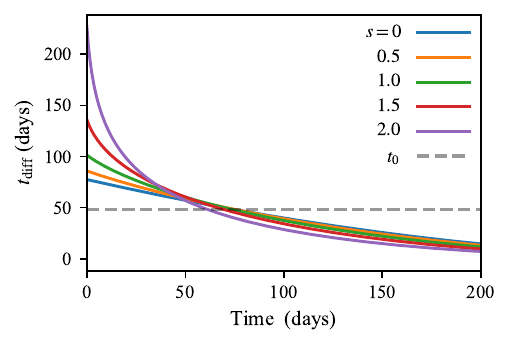}
    \caption{The effective diffusion time $t_{\mathrm{diff}}$ 
between the shock and the photosphere, shown as a function of time. Each colored line represents a different value for the CSM  density slope $s$. The gray dashed line marks the constant timescale $t_0$ assumed by \cite{Chatzopoulos2012}, included for comparison.}
    \label{fig:diffusion_time}
\end{figure}

As shown in Figure ~\ref{fig:diffusion_time} the effective diffusion time between the shock and the photosphere, \(t_{\rm diff}(t)\), is jointly determined by the shock radius (setting the geometric separation) and by the CSM density profile and mean opacity \(\kappa\). As the interaction proceeds, photon path lengths shorten and the effective optical depth decreases, yielding a monotonic decline of \(t_{\rm diff}\) with time. The colored curves denote different CSM density slopes \(s\); larger \(s\) implies a more centrally concentrated (denser) inner CSM and therefore a longer diffusion time at early epochs. At later times, the models converge as the shock expands outward and the photosphere recedes. The gray dashed line marks the constant diffusion timescale \(t_{0}\) commonly assumed in analytic treatments \citep[e.g.,][]{Chatzopoulos2012} and is shown for reference.

\subsection{Light-Curve Morphology and Stages} 
At early times, when the shock radius $R_{\mathrm{sh}}$ lies near the inner CSM boundary, energy is deposited at a rate $L_{\mathrm{heat}}(t)$ but remains trapped at large optical depth. This stage produces little  emission—the ``dark phase''.  As the shocks propagate outward, diffusion gradually transports energy through the unshocked CSM. When the FS reaches layers of lower optical depth, photons begin to escape ahead of the shock, and the luminosity rises toward a peak. At this stage, the emission tracks the instantaneous shock power. Once the FS reaches the outer edge of the CSM, shock heating ceases and the luminosity drops rapidly. Residual photons generated in deeper regions continue to diffuse outward, powering the declining tail of the light curve.

Initially, shocks are near the inner CSM radius (\(r \approx R_\mathrm{csm,in}\)), the relevant diffusion time corresponds to photon escape from this inner region to the outer boundary:
\begin{equation} \label{eq:tdiff_max}
   t_{\mathrm{diff,max}} = \frac{\tau_0 R_{\mathrm{csm,out}} \Theta_{\mathrm{in}}}{c} \approx 17.6\,\mathrm{days} \, \kappa_{0.2} M_{\mathrm{csm},\odot} R_{\mathrm{csm,out},4}^{-1},
\end{equation}
where \(\Theta_{\mathrm{in}}\) is a dimensionless geometric factor of order unity that depends weakly on the CSM density profile (\(\eta_\mathrm{csm}\)), and we have scaled the parameters as \(\kappa_{0.2} \equiv \kappa / (0.2\,\mathrm{cm^2\,g^{-1}})\), \(M_{\mathrm{csm},\odot} \equiv M_{\mathrm{csm}} / M_\odot\), and \(R_{\mathrm{csm,out},4} \equiv R_{\mathrm{csm,out}} / (10^4 R_\odot)\).

The corresponding shock-crossing time for the ejecta to traverse the CSM is approximately:
\begin{equation} \label{eq:tdyn_max}
   t_{\mathrm{dyn,max}} \approx \frac{R_{\mathrm{csm,out}}}{v_{\mathrm{ej}}} \approx 8\,\mathrm{days} \, M_{\mathrm{ej},\odot}^{1/2} E_{\mathrm{sn},51}^{-1/2} R_{\mathrm{csm,out},4},
\end{equation}
where \(v_{\mathrm{ej}} \approx (2 E_{\mathrm{sn}} / M_{\mathrm{ej}})^{1/2}\) is the characteristic velocity of the ejecta, \(M_{\mathrm{ej},\odot} \equiv M_{\mathrm{ej}} / M_\odot\), and \(E_{\mathrm{sn},51} = E_{\mathrm{sn}} / (10^{51}\,\mathrm{erg})\).

Following \citet{Khatami2024}, the ratio of these timescales defines a key dimensionless parameter, \(\xi\), which governs the qualitative morphology of the light curve:
\begin{equation} \label{eq:xi_def}
   \xi \equiv \frac{t_{\mathrm{diff,max}}}{t_{\mathrm{dyn,max}}} \approx 2.2 \, \kappa_{0.2} M_{\mathrm{csm},\odot} M_{\mathrm{ej},\odot}^{-1/2} E_{\mathrm{sn},51}^{1/2} R_{\mathrm{csm,out},4}^{-2}.
\end{equation}
Systems with \(\xi \gg 1\) correspond to optically thick, ``compact'' CSM where diffusion dominates, while \(\xi \ll 1\) corresponds to optically thinner, ``extended'' CSM where the light curve more closely tracks the shock dynamics.

\begin{figure*}[!t]
    \centering
\includegraphics[width=0.45\textwidth]{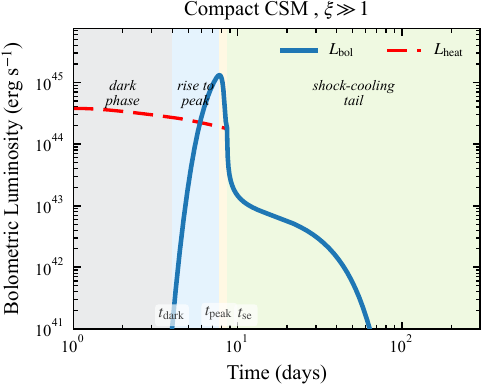}
\includegraphics[width=0.45\textwidth]{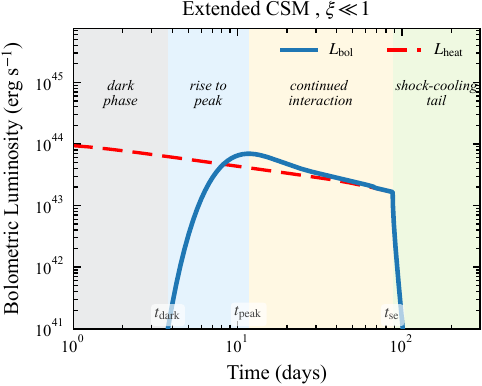}
    \caption{
    Bolometric light-curve morphologies of interaction-powered SNe for compact 
    (\(\xi \gg 1\), left) and extended (\(\xi \ll 1\), right) CSM. 
    The blue lines show the bolometric luminosity \(L_{\rm bol}\), 
    while the red dashed lines indicate the shock power \(L_{\rm heat}\). 
    Vertical dashed lines mark \(t_{\rm dark}\), \(t_{\rm peak}\), and \(t_{\rm se}\). 
    Fixed parameters are 
    \(M_{\rm ej}=5\,M_\odot\), 
    \(E_{\rm sn}=10^{51}\,\mathrm{erg}\), 
    \(M_{\rm csm}=1\,M_\odot\), 
    \(R_{\rm csm,in}=5\times10^{2}\,R_\odot\), 
    \(s=2\), 
    \(\kappa=0.2\,\mathrm{cm^{2}\,g^{-1}}\), 
    and \(\epsilon_{\rm int}=1\). 
    The outer radii are 
    \(R_{\rm csm,out}=5.0\times10^{3}\,R_\odot\) (left) 
    and \(R_{\rm csm,out}=5.0\times10^{4}\,R_\odot\) (right).
    }
    \label{fig:lc_different_xi}
\end{figure*}

Figure \ref{fig:lc_different_xi} illustrates two typical light-curve morphologies for interaction-powered transients, corresponding to interaction with a compact CSM, characterized by a large optical depth parameter (\(\xi \gg 1\), left panel) and an extended CSM (\(\xi \ll 1\), right panel). The evolution of both morphologies can be divided into distinct phases based on the characteristic timescales: \(t_\mathrm{dark}\) (the onset of photon escape), \(t_\mathrm{peak}\) (the time of peak luminosity), and \(t_\mathrm{se}\) (the shock emergence time from the CSM).

\begin{enumerate}
    \item \textbf{Dark Phase (\(t < t_\mathrm{dark}\))}: In this initial phase, although the shock interaction with the CSM has commenced, generating a shock power \(L_\mathrm{heat}\), the shock is deeply embedded within an optically thick medium. Photons are effectively trapped, resulting in an observed bolometric luminosity \(L_\mathrm{bol}\)  that is significantly lower than \(L_\mathrm{heat}\).

    \item \textbf{Rise to Peak (\(t_\mathrm{dark} < t < t_\mathrm{peak}\))}: At \(t \approx t_\mathrm{dark}\), as the shocks moving forward and the optical depth decreases, the photon diffusion time becomes comparable to the dynamical time. Photons begin to escape, and \(L_\mathrm{bol}\) rises rapidly.
        \begin{itemize}
            \item Compact CSM: The long diffusion timescale means the peak luminosity is dominated by the release of thermal energy stored deep within the optically thick material at shock breakout, resulting in a sharp peak.
            \item Extended CSM: The diffusion timescale is relatively shorter. The rise in \(L_\mathrm{bol}\) is more gradual, modulated by both photon diffusion and the concurrently evolving  \(L_\mathrm{heat}\).
        \end{itemize}

    \item \textbf{Post-Peak Evolution (\(t_\mathrm{peak} < t < t_\mathrm{se}\))}: The post-peak evolutionary paths differ significantly:
        \begin{itemize}
            \item Compact CSM: The light curve immediately enters a ``shock-cooling tail'' The decline in \(L_\mathrm{bol}\) reflects the cooling of the post-shock-breakout material due to adiabatic expansion and radiative losses.
            \item Extended CSM: Between \(t_\mathrm{peak}\) and \(t_\mathrm{se}\), the light curve enters a phase of ``continued interaction''. In this regime, the photon diffusion time is much shorter than the evolutionary timescale, allowing the observed luminosity \(L_\mathrm{bol}\) to track the ongoing shock power in near real-time (\(L_\mathrm{bol} \approx L_\mathrm{heat}\)), often presenting as a slowly declining plateau.
        \end{itemize}

    \item \textbf{Shock-Cooling Phase (\(t > t_\mathrm{se}\))}: 
At \(t_\mathrm{se}\), the forward shock exits the outer CSM boundary (\(R_\mathrm{csm,out}\)), and the interaction power \(L_\mathrm{heat}\) terminates. The subsequent luminosity is then sustained only by the residual thermal energy.

\begin{itemize}
    \item Compact CSM: Since \(t_\mathrm{se}\) occurs relatively late, a significant fraction of internal energy remains. The cooling tail is therefore extended, and the decline of \(L_\mathrm{bol}\) is comparatively gradual.
    \item Extended CSM: Most of the internal energy has already been radiated away before shock breakout. Once \(L_\mathrm{heat}\) vanishes at \(t_\mathrm{se}\), the light curve rapidly transitions into a steep shock-cooling decline.
\end{itemize}
\end{enumerate}

In summary, the overall morphology of an interaction-powered transients light curve---including the duration of the dark phase, the speed of the rise, the shape of the peak, and whether the post-peak decline is dominated by continued interaction or shock cooling---is governed by the interplay of the characteristic timescales \(t_\mathrm{dark}\), \(t_\mathrm{peak}\), and \(t_\mathrm{se}\). These timescales, in turn, depend critically on the physical properties of both the supernova ejecta and the CSM. Accurately modeling this evolutionary sequence therefore provides a powerful method for constraining the progenitor's properties and mass-loss history from observed light curves.

\subsection{Dependence on properties of CSM}

\begin{figure*}[!t]
    \centering
\includegraphics[width=0.45\textwidth]{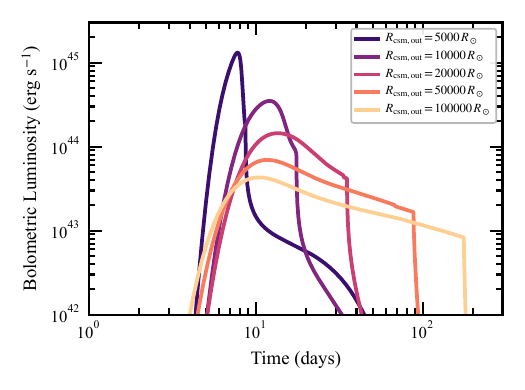}
\includegraphics[width=0.45\textwidth]{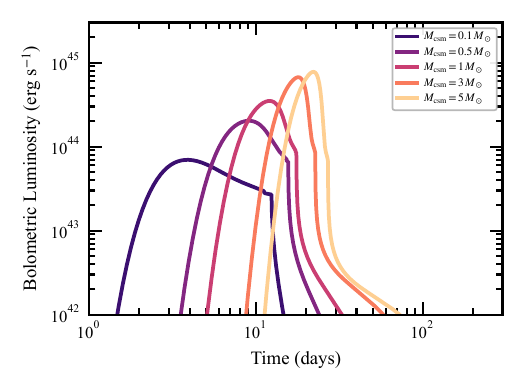}
    \caption{Effect of the CSM outer radius $R_{\rm csm,out}$ (left) and CSM mass $M_{\rm csm}$ (right) on supernova bolometric light curves. In the left panel, $R_{\rm csm,out}$ is varied with $M_{\rm csm}=1M_\odot$; in the right panel, $M_{\rm csm}$ is varied with $R_{\rm csm,out}=1.0\times10^{4}R_\odot$. All other parameters are fixed to $M_{\rm ej}=5M_\odot$, $E_{\rm sn}=10^{51}\mathrm{erg}$, $R_{\rm csm,in}=5.0\times10^{2},R_\odot$, $s=2$, $\kappa=0.2\mathrm{cm^{2}g^{-1}}$, and $\epsilon_{\rm int}=1$.
}
    \label{Fig:lc_different_R_and_M}
\end{figure*}

The morphology of interaction-powered supernova light curves is highly sensitive to the properties of CSM. Variations in the spatial extent, total mass, and density distribution of the CSM regulate the diffusion and heating processes, thereby leaving distinct imprints on both the luminosity scale and temporal evolution. To quantify these dependencies, we examine the roles of $R_{\rm csm,out}$, $M_{\rm csm}$, and the density slope $s$ individually.

The left panel of Figure~\ref{Fig:lc_different_R_and_M} shows the effect of varying the CSM outer radius $R_{\rm csm,out}$ while keeping other parameters fixed. As $R_{\rm csm,out}$ increases from $5\times10^{3}R_\odot$ to $10^{5}R_\odot$, the bolometric light curve exhibits a systematic transition from a compact to an extended CSM regime. The onset of emission is not significantly delayed, indicating that the duration of the dark phase is only weakly sensitive to the CSM extent. Instead, the dominant effect is a progressive broadening of the light curve and a shift of the peak to later times, accompanied by a smoother, less pronounced maximum. This reflects the fact that in more extended CSM, the shock energy is released over a longer timescale, leading to a plateau-like morphology rather than a sharp peak.

While $R_{\rm csm,out}$ mainly determines whether the system falls into a compact or extended regime, the total CSM mass plays an equally crucial role in shaping the light-curve morphology. The right panel of Figure~\ref{Fig:lc_different_R_and_M} presents the effect of increasing the CSM mass $M_{\rm csm}$ at a fixed $R_{\rm csm,out}=10^{4}R_\odot$. A larger CSM mass increases the optical depth and interaction efficiency, which results in a higher peak luminosity. Meanwhile, the dark phase becomes noticeably longer due to the slower photon diffusion, and the light-curve width becomes narrower as the shock-deposited energy is radiated away more efficiently after maximum. This combination of a longer rise and steeper decline produces a more asymmetric light-curve shape.

\begin{figure*}
    \centering
\includegraphics[width=0.45\textwidth]{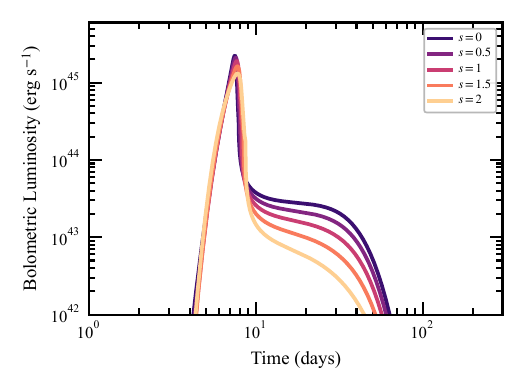}
\includegraphics[width=0.45\textwidth]{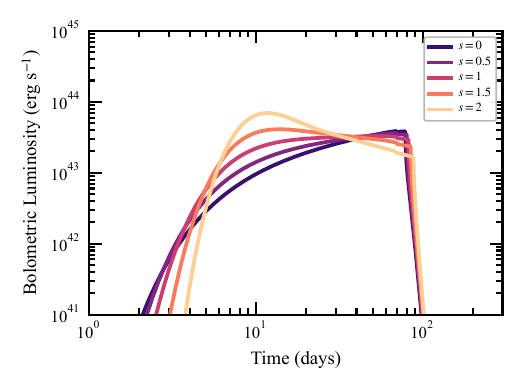}

     \caption{Effect of the CSM density slope $s$ on  bolometric light curves.
The left panel shows the compact CSM case ($R_{\mathrm{csm,out}}=5\times10^{3}R_{\odot}$), where slow photon escape leads to later, sharper peaks and steeper post-peak evolution. 
The right panel shows  the extended CSM case ($R_{\mathrm{csm,out}}=5\times10^{4}R_{\odot}$), where diffusion dominates and produces broader peaks with slower declines.
Other parameters are fixed to $M_{\mathrm{ej}}=5M_{\odot}$, $E_{\mathrm{sn}}=10^{51}$ erg, $R_{\mathrm{csm,in}}=5\times10^{2}R_{\odot}$,  $\kappa=0.2~\mathrm{cm^{2}g^{-1}}$, and $\epsilon_{\mathrm{int}}=1$.}
    \label{Fig:lc_different_s}
\end{figure*}

In addition to the overall extent and mass of the CSM, the density distribution itself can also leave clear imprints on the observed light curve. Figure~\ref{Fig:lc_different_s} illustrates the influence of the CSM density slope $s$ on the morphology of the bolometric light curve. The left panel presents the compact CSM case with $R_{\rm csm,out}=5\times10^{3}R_\odot$. In this regime, the effect of changing the density slope $s$ is modest. Since the photon escape time is short, the overall morphology of the light curves is only weakly sensitive to $s$, and the differences appear mainly in the late-time cooling phase. Steeper density profiles ($s=2$) lead to slightly faster declines after maximum, but the peak timing and luminosity remain nearly unchanged.

By contrast, the right panel shows the extended CSM case with $R_{\rm csm,out}=5\times10^{4}R_\odot$, where the diffusion time is much longer and the role of $s$ becomes significant. Here, the density slope directly regulates the shock heating rate: shallower density distributions provide a more sustained energy input, leading to broader, flatter light curves, while steeper profiles yield later, sharper peaks and a more rapid post-peak decline. The diversity introduced by $s$ in this extended regime highlights the sensitivity of the interaction-powered light curve to the detailed CSM structure.

\subsection{Comparison with Previous Semi-Analytic Models}

\begin{figure}[h]
    \centering
    \includegraphics[width=0.95\linewidth]{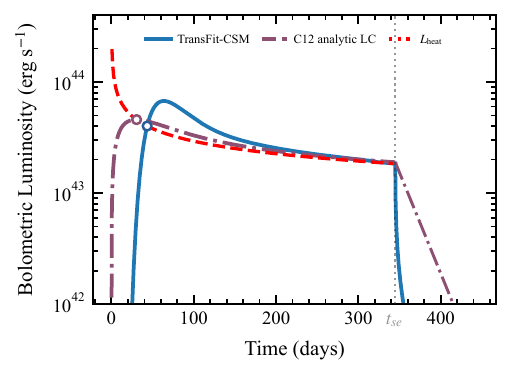}
\caption{Comparison of bolometric light curves computed with the \texttt{TransFit-CSM} model (solid lines) and the semi-analytic prescription of \citet{Chatzopoulos2012} (dashed lines). The C12 treatment, assuming instantaneous and homogeneous energy deposition, naturally satisfies Arnett's rule. In contrast, the \texttt{TransFit-CSM} calculation explicitly includes photon diffusion from a moving shock front. This predicts an early faint phase and a diffusion-mediated peak, demonstrating a breakdown of Arnett's rule when interaction occurs within optically thick CSM.}
    \label{fig:LC_tau}
\end{figure}

Our approach differs in several key respects from the semi-analytic framework of \citet[hereafter C12]{Chatzopoulos2012}, which generalized Arnett’s diffusion formalism to include contributions from both the FS and RS. In their treatment, the shock power is assumed to be centrally deposited within the diffusion mass---a critical simplification that neglects the outward motion of the shocks. In addition, radiative transport is described using a single, fixed effective diffusion time. This central-source, constant-timescale approximation renders the governing equation separable and computationally tractable.

Under these assumptions, the bolometric luminosity takes the form \citep{Chatzopoulos2012}
\begin{equation}
L_{\mathrm{bol}}(t) = \frac{1}{t_0} e^{-t/t_0}
\int_0^t e^{t'/t_0} L_{\mathrm{sh}}(t') \, \mathrm{d}t' ,
\end{equation}
where the effective diffusion timescale is constant,
\begin{equation}
t_0 \equiv \frac{\kappa M_{\mathrm{csm,th}}}{13.8\,c\,R_{\mathrm{ph}}}.
\end{equation}
Here \(t_0\) represents the photon diffusion time through the optically thick CSM mass, evaluated at a fixed photospheric radius. In reality, the photon diffusion time is not fixed; it evolves as the shock propagates outward. Since the effective diffusion path from the shock to the photosphere shortens over time, the escape timescale decreases. A constant \(t_0\) therefore cannot capture the changing efficiency of photon escape. For comparison, \citet{Moriya2013} studied the opposite limit of optically thin CSM, where radiative diffusion is neglected and the emergent luminosity directly follows the instantaneous shock heating rate.For comparison, \citet{Moriya2013} studied the opposite limit of optically thin CSM, where radiative diffusion is neglected and the emergent luminosity directly follows the instantaneous shock heating rate.

Figure~\ref{fig:LC_tau} compares bolometric light curves computed with our \texttt{TransFit-CSM} model and the semi-analytic prescription of \citet[][hereafter C12]{Chatzopoulos2012}. In the \texttt{TransFit-CSM}, the luminosity remains faint during the first few days because the shock is deeply embedded within the optically thick CSM. Once the shock-generated thermal energy diffuses to the outer boundary, the light curve rises sharply, often forming a diffusion-mediated peak. Subsequently, as the shock propagates outward and the optical depth ahead of it decreases (\(\tau \lesssim c/v_\mathrm{sh}\)), the energy deposited at the shock front can escape more rapidly. The light curve may then plateau, with the luminosity more closely tracking the instantaneous shock power. Finally, after the shock breaks out of the CSM, the system enters the cooling phase, and the stored internal energy radiates away, producing a rapid decline.

By construction, the semi-analytic C12 solution assumes instantaneous energy deposition distributed according to the mass profile, which naturally enforces Arnett's rule \citep{Arnett1982}:
\begin{equation} \label{eq:arnett_rule}
   L_\mathrm{peak} \approx L_\mathrm{heat}(t_\mathrm{peak}).
\end{equation}
This rule links the peak luminosity directly to the instantaneous heating rate at the time of the peak. In contrast, our \texttt{TransFit-CSM} calculation explicitly accounts for photon diffusion from a moving source. As Figure~\ref{fig:LC_tau} demonstrates, this reveals that Arnett's rule does not generally hold for interaction with optically thick CSM. The C12 approach consequently misses both the early faint phase and the distinct, diffusion-mediated peak structure captured by \texttt{TransFit-CSM}.

\begin{deluxetable*}{l l c c c c}
\tablecaption{Key Model Parameters and Default Priors for SN~2006gy and SN~2010jl\label{tab:modelparams}}
\tablewidth{0pt}
\tablehead{
  \colhead{Parameter} &
  \colhead{Definition} &
  \colhead{Units} &
  \colhead{Default Prior} &
  \colhead{SN~2006gy} &
  \colhead{SN~2010jl}
}
\startdata
$M_{\mathrm{ej}}$            & Ejecta mass                 & $M_{\odot}$                 & L[$0.1$, $100$]               & $37.8^{+2.16}_{-3.21}$ & $26.6^{+2.3}_{-3.79}$ \\
$E_{\mathrm{sn}}$            & Ejecta kinetic energy       & $10^{51}\,\mathrm{erg}$     & L[$0.1$, $30$]  & $1.57^{+0.46}_{-0.27}$ & $1.1^{+0.13}_{-0.07}$ \\
$M_{\mathrm{csm}}$           & CSM mass                    & $M_{\odot}$                 & L[$0.01$, $80$]         & $46.4^{+2.56}_{-3.72}$ & $10.8^{+1.54}_{-1.71}$ \\
$R_{\mathrm{csm,out}}$       & CSM outer radius            & $10^{4}\,R_{\odot}$         & L[$0.1$, $500$]                & $7.69^{+0.95}_{-0.65}$ & $8.47^{+0.03}_{-0.03}$ \\
$s$                          & CSM density slope           &                       & U[$0$, $3$]                           & $2.66^{+0.03}_{-0.02}$ & $1.67^{+0.01}_{-0.01}$ \\
$\kappa$                     & CSM opacity                 & $\mathrm{cm^{2}\,g^{-1}}$   & U[$0.05$, $0.6$]                      & $0.45^{+0.03}_{-0.05}$ & $0.12^{+0.02}_{-0.01}$ \\
$\epsilon_{\mathrm{int}}$    & Thermalized efficiency      &                        & U[$0.1$, $1.0$]                       & $0.75^{+0.16}_{-0.17}$ & $0.75^{+0.16}_{-0.17}$ \\
$t_{\mathrm{shift}}$         & Explosion time shift        & day                         & U[$0$, $200$]                         & $91.85^{+1.26}_{-1.20}$& $41.1^{+2.49}_{-2.38}$ \\
\enddata
\tablecomments{L = log-uniform prior; U = uniform prior. The \texttt{Default Prior} column lists the priors adopted in our analysis. The inner CSM radius was fixed to $R_{\mathrm{csm,in}}=500\,R_\odot$, ejecta density index $n=10$, and $\delta=1$.}
\end{deluxetable*}

\section{Applications to Observed SNe} \label{sec:Applications}
To demonstrate the practical utility of our framework, we apply the newly developed \texttt{TransFit-CSM} to observed supernovae. This lightweight and modular Python package is designed to fit multi-band and bolometric light curves of interaction-powered transients in a flexible and reproducible way. The code inherits \texttt{TransFit}'s modular data--model--inference workflow and Bayesian machinery, ensuring that observational inputs, physical models, and statistical inference are clearly separated. This allows rapid testing of different prescriptions for CSM structures while preserving numerical accuracy. The mathematical foundations of the fitting procedure remain unchanged; see \citet{Liu2025} for details. Compared to semi-analytical one-zone approaches, \texttt{TransFit-CSM} retains the rigor of explicitly solving the diffusion equation, thereby capturing departures from Arnett-like behavior and more faithfully describing the energy balance in the presence of extended, optically thick CSM.

We then apply \texttt{TransFit-CSM} to two well-observed Type~IIn supernovae. For SN~2006gy, a prototypical interaction-powered event with high-quality coverage, we fit the published bolometric light curve. For SN~2010jl, we perform a joint multi-band fit to the optical photometry, modeling all bands simultaneously with a single parameter set. A comprehensive, sample-level analysis applying \texttt{TransFit-CSM} to a large sample of interaction-powered transients---quantifying parameter distributions and correlations---will be presented in future work.

\subsection{SN 2006gy}

\begin{figure}
    \centering
    \includegraphics[width=1\linewidth]{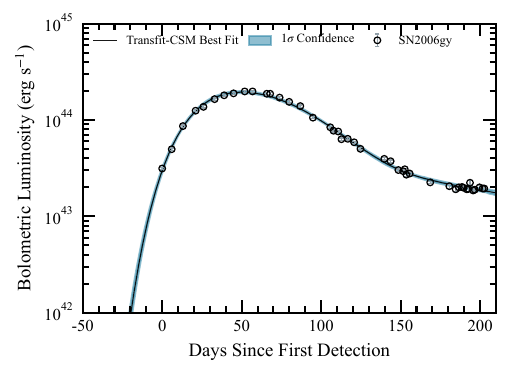}
    \caption{Bolometric light curve of SN 2006gy (black circles) compared with the best-fit \texttt{TransFit-CSM} model (solid line). The points show the observational
data compiled from \cite{Smith2007}. The shaded region indicates the $1\sigma$ confidence interval of the fit.}
    \label{fig:2006gyfitting}
\end{figure}

\begin{figure*}[h]
    \centering
    \includegraphics[width=0.8\linewidth]{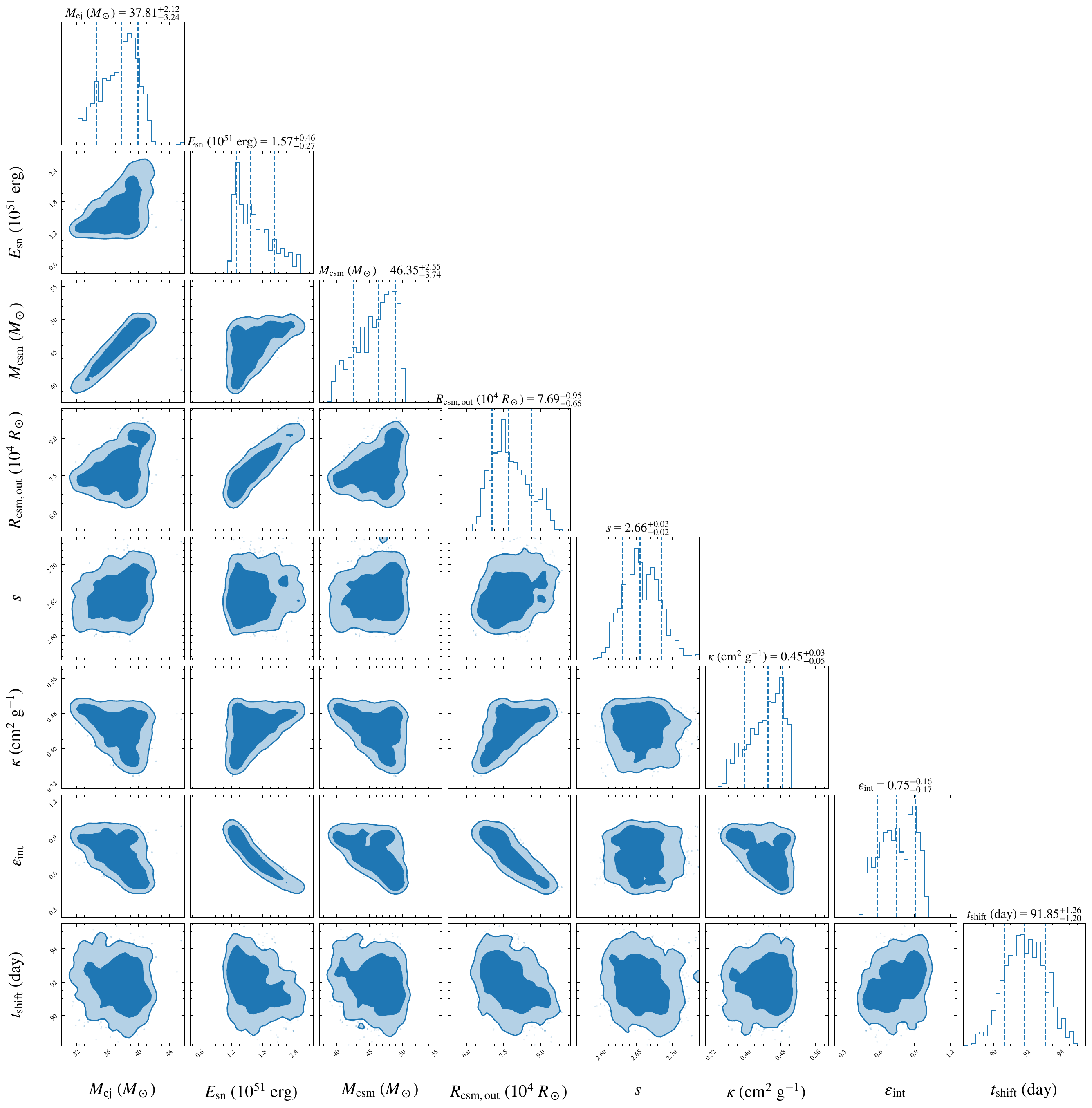}
    \caption{Corner plot showing the posterior probability distributions for the physical parameters of SN 2006gy obtained with our MCMC analysis using TransFit-CSM.}
    \label{fig:SN2006gyconner}
\end{figure*}

SN~2006gy, discovered on 2006 September~18 in the galaxy NGC~1260, is one of the most luminous Type~IIn supernovae ever observed. Its light curve exhibits a slow rise and an extended plateau lasting for several months, characteristic of extremely strong ejecta--CSM interaction. The event has long been considered a benchmark for modeling radiation-dominated shocks and massive pre-explosion mass loss, and its physical parameters have been investigated through a variety of semi-analytic and numerical approaches (e.g., \citealt{Smith2007,Chatzopoulos2013,Moriya2013,Khatami2024}).

Figure~\ref{fig:2006gyfitting} compares the observed bolometric light curve with our best-fit \texttt{TransFit-CSM} model, based on the parameters summarized in Table~\ref{tab:modelparams}. The model successfully reproduces the prolonged rise and broad maximum, as well as the slow decline over \(\gtrsim 100\,\mathrm{days}\), implying a massive CSM surrounding an energetic explosion. The corner plot in Figure~\ref{fig:SN2006gyconner} illustrates the posterior distributions of key parameters derived from our Bayesian inference. We find well-constrained ejecta and CSM masses of \(M_{\mathrm{ej}} = 37.8^{+2.1}_{-3.2}\,M_\odot\) and \(M_{\mathrm{csm}} = 46.35^{+2.55}_{-3.74}\,M_\odot\), together with an explosion energy of \(E_{\mathrm{sn}} = 1.57^{+0.46}_{-0.27} \times 10^{51}\,\mathrm{erg}\) and an extended CSM outer radius of \(R_{\mathrm{csm,out}} = 7.7^{+0.95}_{-0.65} \times 10^{4}R_{\odot}\). The opacity \(\kappa = 0.45^{+0.03}_{-0.05}\,\mathrm{cm^2\,g^{-1}}\) and the conversion efficiency \(\epsilon_{\mathrm{int}} = 0.75^{+0.16}_{-0.17}\) are consistent with moderately optically thick, radiatively efficient interaction.

From the fitted parameters, the resulting dimensionless ratio \(\xi \simeq 0.1\) places SN~2006gy firmly in the ``extended CSM'' regime (\(\xi \ll 1\)). In this regime, the diffusion time is short compared to the dynamical time, allowing photons to escape efficiently. The radiative output therefore closely tracks the instantaneous shock power, producing a broad, interaction-powered plateau rather than a narrow, diffusion-dominated peak. A notable outcome is the inferred CSM density slope of \(s = 2.66^{+0.43}_{-0.35}\), which deviates from the steady wind value (\(s=2\)) and points to a non-steady, time-variable mass-loss episode in the years preceding the explosion.

\subsection{SN 2010jl}

SN~2010jl is a luminous and well-studied Type~IIn supernova, discovered in the irregular galaxy UGC~5189A at a distance of \(\sim 50\,\mathrm{Mpc}\) \citep{Newton2010}. Its classification is confirmed by hallmark spectral features: narrow Balmer emission lines superposed on very broad electron-scattering wings, indicative of sustained ejecta--CSM interaction. This interaction powered an exceptionally bright, slowly declining optical/near-infrared (NIR) light curve for hundreds of days \citep{Zhang2012,Ofek2014,Bevan2020}. At late phases, SN~2010jl also exhibited one of the highest intrinsic H\(\alpha\) luminosities ever observed for a Type~IIn, implying a massive, persistently ionized interaction zone. The presence of a high column-density CSM is further supported by X-ray and radio detections \citep{Chandra2012,Chandra2015}.

\begin{figure}
    \centering
    \includegraphics[width=1\linewidth]{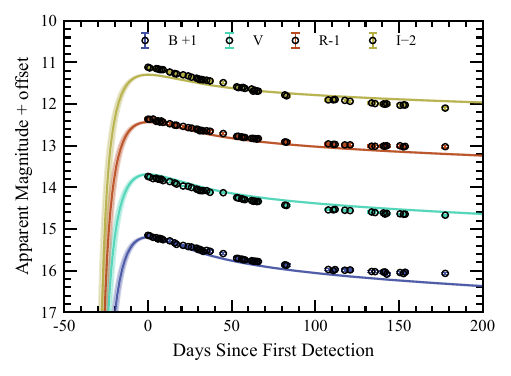}
    \caption{Our \texttt{TransFit-CSM} model fits to the multi-band optical light curves of SN 2010jl. The points show the observational data compiled from \cite{Zhang2012}, while the solid lines and shaded regions represent our best-fit models and their $1 \sigma$ uncertainties, respectively. The light curves are vertically shifted for clarity as noted in the legend.
}
    \label{fig:SN2010jl}
\end{figure}

Figure~\ref{fig:SN2010jl} presents our joint \(B\), \(V\), \(R\), and \(I\)-band fits using the photometry compiled by \citet{Zhang2012}. The maximum-a-posteriori model and its \(1\sigma\) credible region reproduce both the color evolution and the slow decline without band-specific tuning, yielding small and nearly color-independent residuals over \(-20\) to \(+200\,\mathrm{d}\). Our posterior favors \(M_\mathrm{csm} \sim 10\,M_\odot\) with a moderately shallow density slope (\(s \approx 1.7\)), an outer radius of order \( 8\times 10^{4}\,R_\odot\), and an effective thermalization efficiency \(\epsilon_\mathrm{int} \sim 0.7\). Combined with \(M_\mathrm{ej} \sim 27\,M_\odot\) and \(E_\mathrm{sn} \sim 10^{51}\,\mathrm{erg}\), this configuration naturally sustains a radiative shock and explains the persistent luminosity. The implied Thomson optical depth is \(\tau \sim \mathrm{few} \times 10^2\), consistent with the observed electron-scattering wings and the evolution of the X-ray absorbing column \citep{Chandra2012}. These inferences point to extreme, likely eruptive pre-SN mass loss (LBV-like episodes and/or binary interaction) over decades before core collapse \citep{Smith2007}.

\section{Discussions and Conclusions} \label{sec:Conclusions}

In this work, we have developed and applied \texttt{TransFit-CSM}, a new numerical framework for modeling SN--CSM interaction. Unlike previous semi-analytic approaches that rely on separable diffusion equations and central energy deposition, \texttt{TransFit-CSM} explicitly solves the coupled differential equations governing thin-shell shock dynamics and radiative diffusion with a moving, shock-tied heating boundary. This formulation allows us to self-consistently follow the time-dependent evolution of the shock position, optical depth, and photon diffusion time, yielding a physically faithful description of the emergent luminosity.

Our results show that this treatment naturally reproduces the key features of interaction-powered light curves---the early dark phase, the diffusion-mediated rise, the shock-powered peak, and the post-interaction cooling tail. By tracking the outward-propagating shocks, our framework captures the crucial physics inaccessible to traditional analytic prescriptions: the shortening of photon escape paths and the concurrent modification of diffusion timescales. This physical fidelity allows our model to explain the observed diversity of Type~IIn-like transients. We find that the light-curve morphology is governed by the competition between shock heating and this evolving diffusion time, which can be broadly understood in two regimes. A compact CSM produces a pronounced diffusion-dominated peak, followed by a rapid transition into the cooling tail. In contrast, an extended CSM yields a gentler rise and a prolonged, interaction-sustained plateau before cooling sets in.

Our fits to SN~2006gy and SN~2010jl provide clear illustrations of this framework. We find that SN~2006gy, with its broad, slowly evolving plateau, is a classic example of the extended CSM regime, requiring a massive and radially extended envelope. SN~2010jl, while also requiring a substantial CSM reservoir, is fit with a somewhat smaller configuration that still sustains a radiative shock, explaining its persistent, color-stable luminosity.

While our baseline calculations assume a single, spherically symmetric shell, we plan to address more complex progenitor scenarios in the future. For example, very massive stars can undergo pulsational-pair-instability (PPISN) episodes, depositing multiple CSM shells \citep{Renzo2024}, which could explain observed multi-peaked or rebrightening light curves \citep{Liu2018a,Wang2022,Lin2023}. Similarly, a growing set of observations suggests that the CSM is often equatorial or clumpy. A next-generation version of \texttt{TransFit-CSM} could parameterize these departures from spherical symmetry, leveraging multi-wavelength data to disentangle geometry from microphysics \citep[e.g.,][]{Suzuki2019,Kurfurst2020,Wen2024}.

In summary, \texttt{TransFit-CSM} is a  physics-forward Bayesian framework that closes the gap between toy analytic prescriptions and costly radiation-hydrodynamic calculations by explicitly solving the coupled shock dynamics and time-dependent diffusion. By removing ad hoc fixed diffusion times and instantaneous-deposition assumptions, it enables unbiased inference of CSM density profiles, mass-loss histories, and shock energetics. Its efficient, modular implementation supports batch fitting and hierarchical population studies for the large transient samples from ongoing and forthcoming surveys. This enables key applications, including model comparison (interaction versus \(^{56}\mathrm{Ni}\)/magnetar power), real-time follow-up prioritization, and---crucially---a direct route from survey light curves to the physical origin of interaction-powered transients.

\begin{acknowledgements}
We are grateful to Ling-Jun Wang, Zong-Kai Peng for insightful discussions that helped improve this work.
This work is supported by the National Key R\&D Program of China (2021YFA0718500),the National SKA Program of China (2020SKA0120300), the National Natural Science Foundation of China (grant Nos 12303047  and 12393811), Natural Science Foundation of Hubei Province (2023AFB321) and the China Manned Spaced Project (CMS-CSST-2021-A12).
\end{acknowledgements}

\appendix
\section{Dimensionless Formulation of Shock Dynamics and Diffusion}
\label{app:dimensionless}

\texttt{TransFit-CSM} incorporates two coupled components that are solved self-consistently: the dynamics of the thin shell formed by ejecta–CSM interaction, and the diffusion of radiation through the shocked medium. In this appendix, we present the dimensionless formulation and numerical treatment of both aspects.

\subsection{Dynamical Evolution of the Shocked Shell}

\paragraph{Governing equations of dynamical evolution} We approximate the shocked ejecta and CSM as a geometrically thin shell. The dynamics of this shell, described by its radius $R_{\mathrm{sh}}$, velocity $v_{\mathrm{sh}}$, and mass $M_{\mathrm{sh}}$, are governed by the conservation of mass and momentum. The fundamental equations of motion are:
\begin{align}
    \frac{\mathrm{d}R_{\mathrm{sh}}}{\mathrm{d}t} &= v_{\mathrm{sh}}, \\
    \frac{\mathrm{d}M_{\mathrm{sh}}}{\mathrm{d}t} &= 4 \pi R_{\mathrm{sh}}^2 \left[ \rho_{\mathrm{ej}} (v_{\mathrm{ej}} - v_{\mathrm{sh}}) + \rho_{\mathrm{csm}} (v_{\mathrm{sh}} - v_{\mathrm{csm}}) \right], \\
    M_{\mathrm{sh}} \frac{\mathrm{d}v_{\mathrm{sh}}}{\mathrm{d}t} &= 4 \pi R_{\mathrm{sh}}^2 \left[ \rho_{\mathrm{ej}} (v_{\mathrm{ej}} - v_{\mathrm{sh}})^2 - \rho_{\mathrm{csm}} (v_{\mathrm{sh}} - v_{\mathrm{csm}})^2 \right].
\end{align}
Here, $\rho_{\mathrm{ej}}$ and $\rho_{\mathrm{csm}}$ are the densities of the unshocked ejecta and CSM at the shell's location, respectively. The ejecta velocity at the shell is given by $v_{\mathrm{ej}} = R_{\mathrm{sh}}/t$ (assuming homologous expansion of unshocked ejecta), and $v_{\mathrm{csm}}$ is the pre-shock CSM velocity. We adopt the approximation $v_{\mathrm{sh}} \gg v_{\mathrm{csm}}$.

\paragraph{Dimensionless variables.}

We define three characteristic scales: the inner CSM radius, $R_{\mathrm{csm,in}}$; the maximum ejecta velocity, $v_{\mathrm{ej,max}}$; and the resulting timescale, $t_{\mathrm{in}} \equiv R_{\mathrm{csm,in}} / v_{\mathrm{ej,max}}$. The primary physical variables are then scaled as follows:
\begin{equation}
    x \equiv \frac{R_{\mathrm{sh}}}{R_{\mathrm{csm,in}}}, \quad
    w \equiv \frac{v_{\mathrm{sh}}}{v_{\mathrm{ej,max}}}, \quad
    \zeta \equiv \frac{t}{t_{\mathrm{in}}}.
\end{equation}
The density profiles are written in a separable, dimensionless form:
\begin{equation}
    \rho_{\mathrm{ej}}(x,\zeta) = \rho_{\mathrm{ej,in}}\, \zeta^{-3}\,\eta_{\mathrm{ej}}(x,\zeta),
    \qquad
    \rho_{\mathrm{csm}}(x) = \rho_{\mathrm{csm,in}}\, \eta_{\mathrm{csm}}(x),
\end{equation}
where $\eta_{\mathrm{ej}}$ and $\eta_{\mathrm{csm}}$ are dimensionless functions describing the spatial structure of the ejecta and CSM. Finally, we define a dimensionless shell mass:
\begin{equation}
    \phi \equiv \frac{M_{\mathrm{sh}}}{4 \pi R_{\mathrm{csm,in}}^3 \rho_{\mathrm{ej,in}}}.
\end{equation}

\paragraph{Dimensionless dynamics.} Substituting these variables into the governing equations yields a coupled system of first-order ordinary differential equations (ODEs):
\begin{align}
    \frac{\mathrm{d}x}{\mathrm{d}\zeta} &= w, \label{eq:dimless_x} \\
    \frac{\mathrm{d}\phi}{\mathrm{d}\zeta} &= x^2 \zeta^{-3} \left( \frac{x}{\zeta} - w \right)\eta_{\mathrm{ej}}(x,\zeta) + q\, x^2 w\, \eta_{\mathrm{csm}}(x), \label{eq:dimless_phi} \\
    \frac{\mathrm{d}w}{\mathrm{d}\zeta} &= \frac{1}{\phi} \left[ x^2 \zeta^{-3}\left( \frac{x}{\zeta} - w \right)^2 \eta_{\mathrm{ej}}(x,\zeta) - q\, x^2 w^2\, \eta_{\mathrm{csm}}(x)\right], \label{eq:dimless_w}
\end{align}
where $q \equiv \rho_{\mathrm{csm,in}} / \rho_{\mathrm{ej,in}}$ is the characteristic density ratio between the CSM and ejecta.

At the onset of interaction ($\zeta = 1$) we take
\begin{equation}
  x (1) = 1, \qquad w (1) \simeq 1, \qquad \phi (1) \ll 1,
\end{equation}
meaning the fastest ejecta just reach $R_{\textrm{csm,in}}$ with negligible
swept-up mass. The shock-emergence time $\zeta_{\mathrm{se}}$ is defined by $x
(\zeta_{\mathrm{se}}) = R_{\textrm{csm,out}} / R_{\textrm{csm,in}}$; for
$\zeta > \zeta_{\mathrm{se}}$ the shocked CSM freely expands.

\subsection{Interaction Phase (Diffusion with Heating)}

\paragraph{Governing equation.}
During the interaction phase, the energy conservation equation per unit mass is
\begin{equation}
  \frac{\partial E(r,t)}{\partial t}
  \;=\;
  -\,\frac{\partial L}{\partial m}
  \;+\;
  \dot{\epsilon}_{\rm sh}(r,t),
\end{equation}
where $E=u/\rho$ is the specific internal energy, $u(r,t)$ is the radiation energy density, and $\dot{\epsilon}_{\rm sh}(r,t)$ is the specific shock-heating rate. Because the CSM is stationary ($\partial\rho/\partial t=0$),
\begin{equation}
  \frac{1}{\rho}\frac{\partial u}{\partial t}
  \;=\;
  \frac{1}{r^{2}\rho}\frac{\partial}{\partial r}
  \!\left[\frac{c\,r^{2}}{3\kappa\rho}\frac{\partial u}{\partial r}\right].
\end{equation}

\paragraph{Non-dimensionalization.}
Introduce the dimensionless energy density $e(x,t)$ via
\begin{equation}
  u(r,t)=u_{0}\,e(x,t),\qquad x\equiv\frac{r}{R_{\rm csm,in}},
\end{equation}
which yields
\begin{equation}
  \frac{\partial e}{\partial t}
  \;=\;
  \left(\frac{c}{3\kappa\rho_{\rm csm,in}R_{\rm csm,in}^{2}}\right)
  \frac{1}{x^{2}}\frac{\partial}{\partial x}
  \!\left[\frac{x^{2}}{\eta_{\rm csm}(x)}\frac{\partial e}{\partial x}\right].
\end{equation}
The prefactor defines the diffusion time,
\begin{equation}
  t_{\rm diff}\;\equiv\;\left(\frac{c}{3\kappa\rho_{\rm csm,in}R_{\rm csm,in}^{2}}\right)^{-1}
  \;=\;\frac{\tau_{\rm csm,in}\,R_{\rm csm,in}}{c},
  \qquad
  \tau_{\rm csm,in}\equiv 3\kappa\rho_{\rm csm,in}R_{\rm csm,in}.
\end{equation}
We evolve the shell dynamics in $\zeta\equiv t/t_{\rm in}$, and the diffusion in
\begin{equation}
  y \;\equiv\; \frac{t}{t_{\rm diff}},
  \qquad\text{so that}\qquad
  t=\zeta\,t_{\rm in}=y\,t_{\rm diff}\;\;\Longrightarrow\;\; y=\frac{t_{\rm in}}{t_{\rm diff}}\,\zeta.
\end{equation}
In terms of $y$, the diffusion equation takes the compact form
\begin{equation}
  \frac{\partial e}{\partial y}
  \;=\;
  \frac{1}{x^{2}}\frac{\partial}{\partial x}
  \!\left[ D(x)\,\frac{\partial e}{\partial x}\right],
  \qquad
  D(x)\equiv \frac{x^{2}}{\eta_{\rm csm}(x)},
  \label{eq:diffusion_dimless}
\end{equation}
where $D(x)$ reflects the assumed CSM density/opacity profiles.

\paragraph{Inner boundary condition.}
At the moving inner boundary $x_{\rm sh}(y)$ we impose a flux-injection condition:
\begin{equation}
  F \;=\; -\,\frac{c}{3\kappa\rho}\left.\frac{\partial u}{\partial r}\right|_{x=x_{\rm sh}}
  \;=\; \frac{L_{\rm sh}(t)}{4\pi R_{\rm sh}^{2}}
  \;\approx\; \frac{\epsilon}{2}\,\rho_{\rm csm}\,v_{\rm sh}^{3},
\end{equation}
where $\epsilon$ is the radiative efficiency of the shock power (it may be set to unity or absorbed into $u_{0}$ if desired). In dimensionless variables this gives
\begin{equation}
  \left.\frac{\partial e}{\partial x}\right|_{x=x_{\rm sh}}
  \;=\; f_{\rm ib}(y),
  \qquad
  f_{\rm ib}(y)\;=\;\epsilon\,\eta_{\rm csm}^{2}\!\bigl(x_{\rm sh}(y)\bigr)\,w^{3}(y),
\end{equation}
provided we choose, for convenience,
\begin{equation}
  u_{0}\;\equiv\;\frac{\tau_{\rm csm,in}\,v_{\rm ej,max}}{c}
  \left(\tfrac{1}{2}\,\rho_{\rm csm,in}\,v_{\rm ej,max}^{2}\right),
\end{equation}
so that the geometry/opacity factors reduce to the simple $\eta_{\rm csm}^{2}w^{3}$ form.

\paragraph{Outer boundary condition.}
At the CSM surface $x=x_{\rm ph}$ we adopt the Eddington plane–parallel boundary condition. For $\tau=2/3$, $T(x_{\rm ph},y)=T_{\rm eff}(y)$. Assuming radiation-pressure dominance ($u=aT^{4}$), the boundary can be written in flux form,
\begin{equation}
  u(x_{\rm ph},y)=-\frac{4}{3\kappa\rho_{\rm csm}}
  \left.\frac{\partial u}{\partial r}\right|_{r=R_{\rm ph}},
\end{equation}
which, in dimensionless variables, becomes
\begin{equation}
  e(x_{\rm ph},y)=f_{\rm ob}\,
  \left.\frac{\partial e}{\partial x}\right|_{x=x_{\rm ph}},
  \qquad
  f_{\rm ob}\equiv-\frac{4}{\tau_{\rm csm,in}\,\eta_{\rm csm}(x_{\rm ph})}.
\end{equation}

\paragraph{Initial condition.}
The function $f_{\rm init}(x)$ specifies the initial radiation-energy distribution. We adopt a cold CSM prior to breakout, $f_{\rm init}(x)=0$, though pre-heated profiles can be accommodated.

\paragraph{System summary.}
The full system of the diffusion process during the interaction phase can be summarized as 
\begin{equation}
\begin{cases}
\displaystyle
\frac{\partial e(x,y)}{\partial y}
= \frac{1}{x^{2}}\,
\frac{\partial}{\partial x}\!\left[D(x)\,\frac{\partial e(x,y)}{\partial x}\right],
& x_{\rm sh}(y)<x<x_{\rm ph},\;\; y_{\rm in}<y\le y_{\rm se},\\[0.9ex]
\displaystyle
e(x_{\rm ph},y)=f_{\rm ob}\,
\left.\frac{\partial e}{\partial x}\right|_{x_{\rm ph}},
& y_{\rm in}<y\le y_{\rm se},\\[0.9ex]
\displaystyle
\left.\frac{\partial e}{\partial x}\right|_{x_{\rm sh}(y)}=f_{\rm ib}(y),
& y_{\rm in}<y\le y_{\rm se},\\[0.9ex]
\displaystyle
e\!\left(x,\,y_{\rm in}\right)=f_{\rm init}(x),
& x_{\rm sh}\!\left(y_{\rm in}\right)\le x\le x_{\rm ph},
\end{cases}
\label{eq:diffusion_system}
\end{equation}
with \(y_{\rm se}\equiv t_{\rm se}/t_{\rm diff}\) and \(y_{\rm in}\equiv t_{\rm in}/t_{\rm diff}\). The
moving shock boundary $x_{\mathrm{sh}} (y)$ is updated at each step.

\subsection{Shock–Cooling Phase (Expansion without Heating)}
\label{subsec:sc_phase}

After shock emergence at $t=t_{\rm se}$ (i.e., $y=y_{\rm se}$), shock heating
ceases and the shocked CSM expands homologously with constant velocity
$v_{\rm se}$. Let $R_{0}\equiv R_{\rm in}(t_{\rm se})$; for $t\ge t_{\rm se}$
we write
\begin{equation}
  R_{\rm in}(t)=R_{0}+v_{\rm se}\,t .
\end{equation}
With the comoving coordinate $x\equiv r/R_{\rm in}$, the thermodynamic scalings are
\begin{align}
  u(r,t) &= u_{0}\,e(x,t)\left(\frac{R_{0}}{R_{\rm in}}\right)^{4}, \\
  \rho(r,t) &= \rho_{0}\left(\frac{R_{0}}{R_{\rm in}}\right)^{3}\eta_{\rm csm}(x).
\end{align}
The diffusion equation becomes
\begin{equation}
  \frac{\partial e}{\partial y}
  = \frac{1}{x^{2}}\frac{\partial}{\partial x}
    \left[D(x,y)\frac{\partial e}{\partial x}\right],
\end{equation}
with time-dependent diffusion coefficient as
\begin{equation}
  D(x,y)=\frac{x^{2}}{\eta_{\rm csm}(x)}\left(\frac{R_{\rm in}}{R_{0}}\right).
\end{equation}

\paragraph{Inner boundary.}
Heating vanishes; impose an adiabatic condition
\begin{equation}
  \left.\frac{\partial e}{\partial x}\right|_{x_{\min}}=0.
\end{equation}

\paragraph{Outer boundary.}
At $x=x_{\rm ph}$, adopt the Eddington (plane–parallel) closure,
\begin{equation}
  e(x_{\rm ph},y)=f_{\rm ob}(y)
  \left.\frac{\partial e}{\partial x}\right|_{x_{\rm ph}}, \qquad
  f_{\rm ob}(y)=-\frac{2}{\tau_{\rm csm,in}\,\eta_{\rm csm}(x_{\rm ph})}
  \left(\frac{R_{\rm in}}{R_{0}}\right)^{2}.
\end{equation}

\paragraph{Initial condition.}
We treat this epoch as the initial condition of the shock-cooling phase, with the distribution 
of internal energy at breakout provided by the cumulative shock heating that occurred during 
the interaction stage.  

\paragraph{Summary.}
The shock–cooling system is
\begin{equation}
  \left\{
  \begin{array}{ll}
    \displaystyle
    \dfrac{\partial e(x,y)}{\partial y}
    = \dfrac{1}{x^{2}}\dfrac{\partial}{\partial x}
      \!\left[D(x,y)\dfrac{\partial e(x,y)}{\partial x}\right],
    & x_{\min}<x<x_{\rm ph},\;\; y>y_{\rm se}, \\[1.1ex]
    \displaystyle
    e(x,y)=f_{\rm ob}(y)\,\dfrac{\partial e(x,y)}{\partial x},
    & x=x_{\rm ph},\;\; y>y_{\rm se}, \\[1.1ex]
    \displaystyle
    \dfrac{\partial e(x,y)}{\partial x}=0,
    & x=x_{\min},\;\; y>y_{\rm se}, \\[1.1ex]
    \displaystyle
    e(x,y_{\rm se})=e_{\rm int}(x),
    & x_{\min}\le x\le x_{\rm ph},\;\; y=y_{\rm se}.
  \end{array}
  \right.
  \label{eq:system_cooling}
\end{equation}
The cooling-phase luminosity is computed from the surface solution as in the
interaction phase.

\subsection{Numerical Scheme}

The coupled ODEs governing the shell dynamics and the PDEs describing
radiative diffusion are solved simultaneously in a modular framework. The
shell equations, which evolve the shock radius, velocity, and accumulated
mass, are integrated with explicit ODE solvers (e.g., Runge--Kutta schemes),
allowing for efficient time stepping with adaptive step-size control.  

The diffusion system, which being a stiff parabolic equation, is advanced using
implicit finite-difference discretizations. In particular, we adopt the
Crank--Nicolson scheme, which is second-order accurate in both space and time
and unconditionally stable. The discretized system leads to a tridiagonal  matrix equation at each time step, which is efficiently solved using
standard linear algebra routines.  

Boundary conditions are implemented consistently at the inner and outer
radiation fronts, ensuring proper flux conservation across the shock and
allowing for energy leakage at the outer boundary. The initial conditions are
set by mapping the deposited shock energy onto the diffusion grid at the onset
of the interaction phase.  

To ensure numerical stability and accuracy, the ODE and PDE solvers are
coupled via operator splitting: the hydrodynamic shell variables are advanced
first, and the updated shock position and heating terms are then supplied to
the diffusion solver. This approach provides a robust and computationally
efficient method for modeling the coupled shock--diffusion system.

\bibliography{sample701}{}
\bibliographystyle{aasjournalv7}

\end{document}